\documentclass[aps,prb,twocolumn,groupedaddress,amsmath]{revtex4}
\usepackage{graphicx}

\begin{document}

%

\title{$Q$-dependence of the inelastic neutron scattering cross section for molecular spin clusters with high molecular symmetry}

\author{
O. Waldmann } \email[Presently at: Department of Physics, The Ohio State University, Columbus OH43210.\\E-mail:
]{waldmann@physik.uni-erlangen.de}

\affiliation{ Physikalisches Institut III, Universit\"at Erlangen-N\"urnberg, D-91058 Erlangen, Germany}

\date{\today}

\begin{abstract}
For powder samples of polynuclear metal complexes the dependence of the inelastic neutron scattering intensity
on the momentum transfer $Q$ is known to be described by a combination of so called interference terms. They
reflect the interplay between the geometrical structure of the compound and the spatial properties of the wave
functions involved in the transition. In this work, it is shown that the $Q$-dependence is strongly interrelated
with the molecular symmetry of molecular nanomagnets, and, if the molecular symmetry is high enough, is actually
completely determined by it. A general formalism connecting spatial symmetry and interference terms is
developed. The arguments are detailed for cyclic spin clusters, as experimentally realized by e.g. the
octanuclear molecular wheel Cr$_8$, and the star like tetranuclear cluster Fe$_4$.
\end{abstract}


\maketitle

%

\section{Introduction}

Molecular nanomagnets, inorganic complexes with tens of magnetic metal centers linked by organic ligands, are a
new class of magnetic materials exhibiting fascinating quantum properties. For instance, quantum tunneling of
the magnetization has been observed in clusters like Mn$_{12}$ and  Fe$_8$,\cite{Mn12_Fe8} now called single
molecule magnets. Another class of molecular nanomagnets is established by the molecular wheels as represented
by the prototype Fe$_{10}$.\cite{Fe10} These antiferromagnetic cyclic clusters show pronounced steps in the
magnetization curve at low temperature signaling quantum size effects.\cite{Fe10,FWs} Furthermore, the
possibility of coherent tunneling of the N\'eel-vector has been suggested in these compounds.\cite{AC_NVT} Yet
other systems like molecular grids,\cite{OW_Co2x2,OW_Mn3x3,LK_gridsrev1,LK_gridsrev2} tetranuclear star like
clusters,\cite{Fe4} the cluster V$_{15}$,\cite{IC_V15} or the magnetic Keplerate Mo$_{72}$Fe$_{30}$
\cite{JS_Fe30} attracted much interest due to their peculiar magnetic properties.

The magnetism of these polynuclear complexes is, in principle, described by a microscopic spin Hamiltonian which
to a very good first approximation consists of terms representing isotropic Heisenberg exchange interactions and
dipole-dipole interactions among different spin centers within a molecule, and terms due to the ligand field and
Zeeman interaction for each single ion.\cite{OK,AB_DG} Intermolecular interactions are usually very small and
can be neglected. The isotropic Heisenberg terms are dominant in general and these clusters may be thus
described as (antiferromagnetic) Heisenberg spin clusters with weak magnetic anisotropy.

For a detailed understanding of the properties of molecular nanomagnets it is of great importance to be able to
study both the effects due to the Heisenberg interaction and anisotropic terms. Although rarely achieved in
praxis for these rather large molecules, the ultimate goal would be to determine all relevant parameters of the
microscopic spin Hamiltonian like exchange constants, single ion zero-field-splittings, etc.. Besides the
obvious relevance in understanding the physics, the results are of great help also for synthetic chemists as
they allow to establish magneto-structural correlations which in turn provide ideas for a controlled improvement
of these systems.\cite{OK}

Inelastic neutron scattering (INS) has been demonstrated to be a powerful experimental tool in this regard.
\cite{AF_ins,HUG_revins,RC_Fe8ins,LM_Mn12ins} Being a spectroscopic technique, INS provides direct access to
energy splittings. The INS cross section is controlled by the favorable selection rule $\Delta S = 0,\pm1$. INS
thus allows to detect the splitting of individual spin multiplets, as EPR ($\Delta S =0 $), but also the
splitting produced by the magnetic interactions. As a result, INS provides a very straightforward determination
of exchange parameters. It requires, however, large amounts of sample and, frequently, deuteration. Accordingly,
only powder samples of molecular nanomagnets were investigated so far.

As a unique feature, INS additionally allows to measure the dependence of the INS intensity on the scattering
vector ${\bf Q}$. In this work only the situation of powder samples will be considered, i.e. the INS cross
section has to be averaged in ${\bf Q}$ space and the scattering intensity will depend only on the momentum
transfer $Q$.\cite{AF_ins,HUG_revins} On the one hand, the $Q$-dependence allows to distinguish magnetic
excitations from vibrational excitations.\cite{HUG_Qvibrational} On the other hand, and more importantly, the
$Q$-dependence allows discriminating clearly between various types of magnetic
transitions.\cite{AF_ins,HUG_revins,HUG_Qassignmet} Thus, the observed $Q$-dependence of a transition can be
used in addition to its energy for a spectroscopic assignment removing ambiguities in many cases. This advantage
has been explored in several works, a particular illustrative example is given by the tetranuclear cluster
[Co$_4$(H$_2$O)$_2$(PW$_9$O$_{34}$)$_2$] for which the $Q$-dependence was key to establish the correct
model.\cite{HA_Co4ins}

As for the origin of the $Q$-dependence, it is well known that it depends sensitively on the wave functions of
the states involved in a transition.\cite{AF_ins,HUG_revins} In the formula for the INS cross section the matrix
elements for individual spin centers are correlated to their geometrical arrangement and the momentum transfer
via a term $\exp\left[i{\bf Q}({\bf R}_i-{\bf R}_j)\right]$. Here, ${\bf R}_i$ denotes the position vector of
the $i$-th metal center. This gives rise to characteristic interference terms in the INS cross section for
powder samples.\cite{AF_interference} This observation was expressed in Ref.~[\onlinecite{HUG_Qassignmet}]
loosely as

\begin{equation}
\label{I_and_interference}
 \bar{I}(Q) \propto F^2(Q)\left[\text{combination of interference terms}\right],
\end{equation}

where $\bar{I}(Q)$ denotes the averaged INS intensity and $F(Q)$ the magnetic form factor. It is thus clear, and
well understood, that the $Q$-dependence basically senses the spatial properties of the involved wave functions.
However, to the best of the authors knowledge, beyond this basic understanding the significance of
Eq.~(\ref{I_and_interference}) remained unexplored for larger spin systems like molecular nanomagnets.

For a limited number of small systems the $Q$-dependence could be calculated
analytically,\cite{AF_ins,HUG_Qassignmet,AF_interference,AS_Ni2ins,AF_Tb2ins,HA_Co2ins} thus detailing
Eq.~(\ref{I_and_interference}) for these cases. For larger, or more complex systems one has to resort to
numerical procedures \cite{HA_Co4ins,JMCJ_Ni4ins,HA_Co5ins,JMCJ_Co3ins} and a general program has been
developed.\cite{JJBA_powderins} However, the "black box" character of this approach did not yield insight into
the relation for large spin systems between geometrical structure and wave functions as expressed by the
interference terms.

This work aims at showing how the $Q$-dependence is influenced by the molecular symmetry of the spin cluster and
in particular, if the molecular symmetry is high enough, that it is actually predetermined by it. The arguments
will be detailed for the particular cases of the cyclic spin clusters with focus on the molecular wheel
Cr$_8$,\cite{JVS_Cr8synthesis} and the tetranuclear star like cluster Fe$_4$.\cite{Fe4} The molecular wheel
Cr$_8$ and star like cluster Fe$_4$ were chosen as examples as they both exhibit a very high molecular symmetry
and detailed INS experiments were performed recently.\cite{SC_Cr8ins,GA_Fe4ins,GA_Fe4interins} Additionally, the
Heisenberg interaction is much stronger than the anisotropic terms in both compounds. INS measurements on powder
samples thus provide an excellent view on their internal spin structure. Several transitions between different
spin multiplets were observed experimentally and only these are of interest here. As a main result it will be
demonstrated that for such highly symmetric systems the $Q$-dependence actually can be used to determine
experimentally the spatial quantum numbers of the states involved. The molecular wheel Cr$_8$ is a particularly
illustrative example in this respect.

In the next section, first the general equation for the INS cross section of powder samples will be derived,
correcting an earlier result.\cite{JJBA_powderins} Then, in section~\ref{sSYM} the general framework based on
standard group theoretical procedures will be developed. In sections~\ref{sCYC} and \ref{sTET} the molecular
wheel Cr$_8$ and the star like cluster Fe$_4$ will be discussed. The work concludes with section~\ref{sCON}.

%

\section{INS cross section for powder samples}
\label{sINS}

The differential neutron scattering cross section is \cite{AF_ins,HUG_revins}

\begin{equation}
\label{cross_section} {d^2 \sigma \over d\Omega d\omega} = C(Q,T) \sum_{nm} {e^{-\beta E_n} \over Z(T)}
I_{nm}({\bf Q}) \delta(\omega-{E_m-E_n\over \hbar})
\end{equation}

where $C(Q,T)=(\gamma e^2/m_e c^2) (k'/k) \exp[-2 W(Q,T)]$,  $\beta = 1/(k_B T)$, $Z(T)$ the partition function,
and

\begin{eqnarray}
\label{I_nm} I_{nm}({\bf Q}) = \sum_{ij} F^*_i(Q) F_j(Q) e^{i{\bf Q}\cdot{\bf R}_{ij}} \times\cr
\sum_{\alpha\beta} ( \delta_{\alpha\beta}-{Q_{\alpha}Q_{\beta} \over Q^2} ) \langle n|S_{i\alpha}|m\rangle
\langle m|S_{j\beta}|n\rangle.
\end{eqnarray}

In this equation, $F_i(Q)$ is the magnetic form factor of the $i$-th spin center, $\alpha,\beta = x,y,z$, ${\bf
Q} = {\bf k}' - {\bf k}$ is the transferred momentum, and ${\bf R}_{ij}={\bf R}_i-{\bf R}_j$ with the spin
position vectors ${\bf R}_i$. In the following, the abbreviations $l_{\alpha} = Q_{\alpha} / Q$ and
$\tilde{S}_{i\alpha}$ for the matrix elements of $S_{i\alpha}$ are used. In an ordered product
$\tilde{S}_{i\alpha} \tilde{S}_{j\beta}$, the first term always has to be understood as a matrix element between
$\langle n|$ and $|m\rangle$ and the second as between $\langle m|$ and $|n\rangle$.

The cross section for a powder sample is obtained by averaging Eq.~(\ref{cross_section}) over all directions of
${\bf Q}$, i.e. one has to calculate the expression

\begin{equation}
\bar{I}_{ij,nm}(Q) = \int {d\Omega \over 4\pi} e^{i{\bf Q}\cdot{\bf R}_{ij}} \sum_{\alpha\beta}
(\delta_{\alpha\beta}-l_{\alpha}l_{\beta}) \tilde{S}_{i\alpha}\tilde{S}_{j\beta}.
\end{equation}

This is conveniently done by resorting to the calculus of irreducible spherical tensors.\cite{AM_grouptheory}
The sum over $\alpha$ and $\beta$ can be rearranged to $\sum_{\alpha} \tilde{S}_{i\alpha}\tilde{S}_{j\alpha} -
(\sum_{\alpha} l_{\alpha}\tilde{S}_{i\alpha}) (\sum_{\beta} l_{\beta}\tilde{S}_{j\beta})$. The first term in
this expression is easily averaged since it does not depend on the orientation of ${\bf Q}$, yielding $j_0(Q
R_{ij}) {\bf \tilde{S}}_i\cdot{\bf \tilde{S}}_j$. Here, $j_k$ is the spherical Bessel function of order $k$. The
second term is calculated by expressing the scalar product $\sum_{\alpha} l_{\alpha}\tilde{S}_{i\alpha}$ by
spherical tensors \cite{AM_grouptheory} according to

\begin{equation}
\label{scalar} \sum_{\alpha} l_{\alpha}\tilde{S}_{i\alpha} = \sum_q T^{(1)*}_q({\bf Q})T^{(1)}_q({\bf
\tilde{S}}_i).
\end{equation}

$T^{(k)}_q({\bf O})$ denotes the $q$-th component of the irreducible spherical tensor of degree $k$ with respect
to the vector ${\bf O}$. They are proportional to the spherical harmonics $Y^{k}_q({\bf O})$. Expanding the
exponential in spherical harmonics and using the coupling rule for spherical tensors one obtains

\begin{eqnarray}
\bar{I}_{ij,nm}(Q) = \sqrt{4\pi} \sum_{L=0,2} i^L j_L(Q R_{ij}) \left(\begin{array}{ccc}
1&1&L\\0&0&0\\\end{array}\right) \times\cr \sum_M Y^{L*}_M({\bf R}_{ij}) \left[T^{(1)}({\bf \tilde{S}}_i)
\otimes T^{(1)}({\bf \tilde{S}}_j)\right]^{(L)}_M.
\end{eqnarray}

Here, $(...)$ denotes a Wigner-3$j$ symbol. The term for $L=0$ gives ${1\over 3} j_0(Q R_{ij}) {\bf
\tilde{S}}_i\cdot{\bf \tilde{S}}_j$. For $\bar{I}_{nm}(Q) = \int d\Omega / 4\pi I_{nm}({\bf Q})$ one finally
obtains

\begin{eqnarray}
\label{I} \bar{I}_{nm}(Q) = \sum_{ij} F^*_i(Q) F_j(Q) ( {2 \over 3} j_0(Q R_{ij}) {\bf \tilde{S}}_i \cdot {\bf
\tilde{S}}_j + \cr j_2(Q R_{ij}) \sum_M T^{(2)*}_M({\bf R}_{ij}) \left[T^{(1)}({\bf \tilde{S}}_i) \otimes
T^{(1)}({\bf \tilde{S}}_j)\right]^{(2)}_M ).
\end{eqnarray}

The $L=2$ term could be put also into the more compact form $j_2(Q R_{ij}) {\bf T}^{(2)}({\bf R}_{ij}) \cdot
{\bf T}^{(2)}({\bf \tilde{S}}_i {\bf \tilde{S}}_j)$ using Eq.~(\ref{scalar}).

Several special cases shall be discussed.  It is interesting to consider the limit $Q \rightarrow 0$. Since
$j_0(x\rightarrow 0) \approx 1+\mathcal{O}(x^2)$ and $j_2(x\rightarrow 0) \approx x^2/5$, one simply obtains

\begin{equation}
\label{I_Qto0} \bar{I}_{nm}(Q\rightarrow 0) = {2 \over 3} \sum_{ij} F^*_i(Q) F_j(Q) {\bf \tilde{S}}_i\cdot{\bf
\tilde{S}}_j + \mathcal{O}(Q^2).
\end{equation}

Next, for an isotropic spin cluster the $L=2$ term in Eq.~(\ref{I}) obviously vanishes. If one further numbers
the different possible values of $R_{ij}$ by $p$ and writes $R_p$, the result assumes the convenient form

\begin{equation}
\label{I_isotropic} \bar{I}^{iso}_{nm}(Q,T) = \sum_{p} I^p_{nm}(Q,T) j_0(Q R_p).
\end{equation}

Accordingly, the scattering intensity is sort of an expansion in $j_0(Q R_p)$. In principle, since the $R_p$ are
known, the coefficients $I_p$ can be determined by fitting the $Q$-dependence of the scattering intensity to
Eq.~(\ref{I_isotropic}). Equation~(\ref{I_isotropic}) precises Eq.~(\ref{I_and_interference}) for the case of
isotropic systems.

Finally, one may consider the case where only the components with $\alpha = \beta$ contribute. This holds if
$(\tilde{S}_{i,\alpha} \tilde{S}_{j\beta} + \tilde{S}_{i\beta} \tilde{S}_{j\alpha}) = 0$ for all $\alpha \neq
\beta$. However, it is clear that this condition is not fulfilled in general. One necessary condition is that
the quantization axis for the spin operators coincide with a magnetic main axis. For uniaxial and isotropic
systems this is sufficient to ensure the above condition. However, for a biaxial magnetic system this is not
sufficient in general since in the neutron scattering cross section the local terms $\tilde{S}_{i\alpha}
\tilde{S}_{j\beta}$ enter while "biaxial" is a property of the whole spin cluster as an entity. Thus, for spin
systems with magnetic symmetry lower than uniaxial the validity of this simplification has to be checked
carefully. For an uniaxial system only the $M=0$ term contributes, resulting in

\begin{eqnarray}
\label{I_uniaxial} \bar{I}^{uni}_{nm}(Q) = {2 \over 3} \sum_{ij} F^*_i(Q) F_j(Q) \times\cr \{ \left[ j_0(Q
R_{ij}) - {1\over 2} j_2(Q R_{ij}) C^2_0({\bf R}_{ij}) \right] \left(
\tilde{S}_{ix}\tilde{S}_{jx}+\tilde{S}_{iy}\tilde{S}_{jy} \right) \cr + \left[ j_0(Q R_{ij})+ j_2(Q R_{ij})
C^2_0({\bf R}_{ij}) \right] \tilde{S}_{iz}\tilde{S}_{jz} \}
\end{eqnarray}

with $C^2_0({\bf R}_{ij}) = \left[3(R_{ij,z}/R_{ij})^2-1\right]/2$.

%

\section{INS cross section and spatial symmetry}
\label{sSYM}

The spatial symmetry of the molecule leads to an invariance of the spin Hamiltonian upon certain permutations of
the spin centers.\cite{MT_grouptheory,OW_symmetry} The group of all these permutations is denoted as
$\mathcal{G}$, and an element of this group as $P$. It should be noted that $\mathcal{G}$ and the point group of
the molecule are not equivalent in general. This symmetry of the spin Hamiltonian is accordingly denoted here as
spin permutational symmetry.\cite{OW_symmetry} Its effects are exploited by applying the standard results of
group theory.\cite{AM_grouptheory,MT_grouptheory}

The eigenstates $|n\rangle$ of the spin Hamiltonian can be classified by the irreducible representations (IRs)
of $\mathcal{G}$ and are written now as $|\tau k \mu\rangle$. They transform according to $O(P) |\tau k
\mu\rangle = \sum_{\bar{\mu}} \Gamma_{\bar{\mu}\mu}^{(k)}(P) |\tau k \bar{\mu}\rangle$. $O(P)$ denotes the
operator associated with the permutation $P$. Furthermore, $O(P) S_{i\alpha} O(P^{-1}) = \Gamma_{ji}(P)
S_{j\alpha}$. Here, no sum over $j$ arises on the r.h.s since $P$ is a permutation, i.e. $Pi \neq Pj$ for $i
\neq j$. Accordingly, each row or column, respectively, of the matrix ${\bf \Gamma}(P)$ has only exactly one
nonzero entry which is equal to 1. With these relationships one confirms

\begin{eqnarray}
\label{Sj_to_Si} \langle \tau' k' \mu'|S_{j\beta}|\tau k \mu\rangle = \cr \sum_{\bar{\mu}' \bar{\mu}} \langle
\tau' k' \bar{\mu}'|S_{i\beta}|\tau k \bar{\mu}\rangle \Gamma_{\bar{\mu}'\mu'}^{(k')*}(P_{ji})
\Gamma_{\bar{\mu}\mu}^{(k)}(P_{ji})
\end{eqnarray}

where $P_{ji}$ is a symmetry element which maps the $j$-th spin exactly onto the $i$-th spin, i.e. $P_{ji}j=i$.
Equation~(\ref{Sj_to_Si}) describes the restrictions imposed by the molecular symmetry on the interference
effects in inelastic neutron scattering, and is thus central to this work.

The spin centers can be divided into classes such that all the spin centers of one class are related by the
permutations of the group $\mathcal{G}$. More precisely, a class is a set of spin centers which transform one
into another under the operations of $\mathcal{G}$. The different classes will be numbered by $\gamma$. By
construction, each spin center is member of exactly one class. Therefore,

\begin{eqnarray}
\label{sumij_to_sumgamma} \sum_{i} = \sum_{\gamma} \sum_{i\in\gamma}.
\end{eqnarray}

From each class, one spin center is chosen (arbitrarily) and called the pivotal center of this class. It will be
also indicated by $\gamma$. Equation~(\ref{Sj_to_Si}) then essentially provides a relation between the matrix
element of an arbitrary spin center with the matrix element of its pivotal center. This imposes a certain
structure on the expression $\sum_{ij} F^*_i(Q) F_j(Q) f({\bf Q},{\bf R}_{ij}) \tilde{S}_{i\alpha}
\tilde{S}_{j\beta}$, which, with $f({\bf Q},{\bf R}_{ij})$ appropriately chosen, is the part of the cross
section relevant here. This idea can be used to work out general expressions based on the generalized
Wigner-Eckart theorem.\cite{AM_grouptheory}

For spin permutational groups with only one-dimensional IRs, writing the eigen states as $|\tau k\rangle$,
Eq.~(\ref{Sj_to_Si}) simplifies to $\langle\tau'k'|S_{j\beta}|\tau k\rangle = \langle\tau'k'|S_{i\beta}|\tau
k\rangle \hat{\chi}^{k'k}(P_{ji})$. Whereby, $\hat{\chi}^{kk'}(P) = \chi^{k*}(P) \chi^{k'}(P)$ and $\chi^{k}$
being the character of the $k$-th IR. Together with Eq.~(\ref{sumij_to_sumgamma}) and after some rearrangement
of terms one obtains the central result of this work:

\begin{subequations}
\label{main_result}
\begin{eqnarray}
\label{ggf}
 \sum_{ij} F^*_i(Q) F_j(Q) f({\bf Q},{\bf R}_{ij}) \tilde{S}_{i\alpha} \tilde{S}_{j\beta} = \cr
\sum_{\gamma\bar{\gamma}} \tilde{S}_{{\gamma}\alpha} \tilde{S}_{\bar{\gamma}\beta} F^*_\gamma(Q)
F_{\bar{\gamma}}(Q) f^{kk'}_{\gamma\bar{\gamma}}({\bf Q})
\\
\label{f}
 f^{kk'}_{\gamma\bar{\gamma}}({\bf Q}) =  \sum_{i\in\gamma}\sum_{j\in\bar{\gamma}} f({\bf Q},{\bf
R}_{ij}) \hat{\chi}^{kk'}(P_{i\gamma}) \hat{\chi}^{k'k}(P_{j\bar{\gamma}}).
\end{eqnarray}
\end{subequations}

The $Q$-dependence of the neutron scattering cross section is given by the interference terms
$f^{kk'}_{\gamma\bar{\gamma}}({\bf Q})$ which can be calculated without knowledge of the wave functions, i.e.
are completely governed by the spatial symmetry properties. These equations, being a precise statement of
Eq.~(\ref{I_and_interference}), become the more useful the smaller the number of classes $\gamma$ produced by
the spin permutational symmetry is.

%

\section{Cyclic spin clusters}
\label{sCYC}

The full spin permutational symmetry group of a cyclic spin cluster is $D_N$, where $N$ is the number of spin
centers. However, the spin levels may be also classified by the IRs of the subgroup $C_N$ reflecting the
translational invariance.\cite{MT_grouptheory} This effectively introduces the shift quantum number
$q=0,\ldots,N-1$ via the shift operator $T$: $T|\tau q\rangle = e^{-iq2\pi/N}|\tau q\rangle$. The group elements
may be written as $T^n$ with $n=0,\ldots,N-1$ and the characters are $\chi^{(q)}(T^n) = e^{-iqn2\pi/N}$. The
full symmetry group $D_N$ has one-dimensional and two-dimensional IRs. Spin levels with $q=0$ and $q=N/2$ belong
to one-dimensional IRs of $D_N$, while for $q\neq0,N/2$ the two states with $q$ and $N-q$ belong to a
two-dimensional IR. The latter are thus degenerate.

The interference terms $f^{kk'}_{\gamma\bar{\gamma}}({\bf Q})$ can now be calculated. For each spin center there
is always one symmetry element $T^n$ which connects it to any other spin center. Thus, there is only one class
$\gamma$: $\gamma = \{1,\ldots,N\}$. Spin center 1 is chosen as pivotal center. For $P_{i\gamma}$ one
establishes that $P_{i\gamma} = P_{i1} = T^{i-1}$, and accordingly that $\hat{\chi}^{qq'}(P_{i\gamma})
\hat{\chi}^{q'q}(P_{j\gamma})= e^{i(q-q')(j-i)2\pi/N}$. Introducing the "distance" $n$ by $j=i+n$ yields

\begin{eqnarray}
f^{qq'}({\bf Q}) = \sum_{n=0}^{N-1} \sum_{i=1}^N f({\bf Q},{\bf R}_{i,i+n}) e^{i{2\pi\over N}(q-q')n} .
\end{eqnarray}

The cyclic spin clusters synthesized so far exhibit at best a molecular $S_N$ symmetry axis, as determined by
x-ray crystallography.\cite{AC_Fe6,RWS_Fe6Fe8,BP_Fe6} The magnetic anisotropy of these systems is thus expected
to be strictly uniaxial, as is supported by experiments.\cite{OW_Fe6ins,OW_CsFe8} The position vectors ${\bf
R}_i$ point to above and below the plane of the molecule perpendicular to the main symmetry axis. However, even
with this effect $f({\bf Q},{\bf R}_{i,i+n})$ becomes independent on $i$ for powder samples, i.e. $R_{i,i+n}
\rightarrow R_n$ and $C^2_0({\bf R}_{i,i+n}) \rightarrow C^2_0({\bf R}_{n})$ [see Eq.~(\ref{I_uniaxial})]. Thus,
for powder samples of these highly symmetric, cyclic spin clusters the interference terms are given by

\begin{eqnarray}
\label{f_cyclic}
 f^{qq'}({\bf Q}) = N \sum_{n=0}^{N-1} f({\bf Q},{\bf R}_n) e^{i{2\pi\over N}(q-q')n}.
\end{eqnarray}

The averaged INS cross section for the transition $|\tau q\rangle \rightarrow |\tau'q'\rangle$ is then simply
proportional to

\begin{subequations}
\label{I_cyclic}
\begin{eqnarray}
\bar{I}_{\tau q\tau'q'}(Q) = {2\over3} N F^2(Q) \{ {\bf \tilde{S}}_1 \cdot {\bf \tilde{S}}_1 f_{qq'}^{0}(Q) +
\cr {1\over2} \left( 2 \tilde{S}_{1z}\tilde{S}_{1z} - \tilde{S}_{1x}\tilde{S}_{1x} -
\tilde{S}_{1y}\tilde{S}_{1y} \right) f_{qq'}^{2}(Q) \}
\end{eqnarray}

with

\begin{eqnarray}
f^{0}_{qq'}(Q) =&& \sum_{n=0}^{N-1} j_0(Q R_n) e^{i{2\pi\over N}(q-q')n} \\
f^{2}_{qq'}(Q) =&& \sum_{n=0}^{N-1} j_2(Q R_n) C^2_0({\bf R}_n) e^{i{2\pi\over N}(q-q')n}.
\end{eqnarray}
\end{subequations}

Apart from two normalizing factors, the $Q$-dependence of the INS intensity of uniaxial cyclic clusters is
analytically determined by their spatial symmetry properties. The interference terms $f^{0}_{qq'}(Q)$ and
$f^{2}_{qq'}(Q)$ do not depend on $\tau$ and $\tau'$; they are completely determined by the spatial quantum
numbers.

The following general result is noteworthy. In the limit $Q \rightarrow 0$, Eq.~(\ref{I_cyclic}) leads to

\begin{eqnarray}
\label{I_cyclic_Qto0}
 \bar{I}_{\tau q\tau'q'}(Q) &=& {2\over3} N F^2(Q) {\bf \tilde{S}}_1^2 \sum_{n=0}^{N-1}
e^{i{2\pi\over N}(q-q')n} + \mathcal{O}(Q^2) \cr &\propto& \delta_{qq'} + \mathcal{O}(Q^2).
\end{eqnarray}

Therefore, comparing with Eq.~(\ref{I_Qto0}), the spatial symmetry of the cyclic spin cluster enforces the INS
intensity to become strictly zero for $Q=0$ if $q \neq q'$, i.e. if the spatial quantum numbers of the states
involved are different. In this case, the intensity actually approaches zero as $I(Q) \propto Q^2$.
Equation~(\ref{I_cyclic_Qto0}) might have important experimental implications. Recently it became an intensively
discussed question whether cyclic clusters like the ferric wheels are accurately described by a spin Hamiltonian
with perfect cyclic symmetry, or whether additional terms with less symmetry are of
importance.\cite{MA_levelcrossings,OW_butterflies} In principle, Eq.~(\ref{I_cyclic_Qto0}) provides a powerful
experimental approach to unravel this question since if symmetry breaking terms are relevant the INS intensity
would not drop to zero with $Q \rightarrow 0$.

Eq.~(\ref{I_cyclic}) shall be worked out explicitly for the case of an octanuclear ring with predominantly
antiferromagnetic next neighbor interactions, as appropriate for the so called Cr$_8$
cluster.\cite{JVS_Cr8synthesis} This is motivated by recent, very detailed and accurate INS measurements on this
system.\cite{SC_Cr8ins} The $Q$-dependence of the INS intensity has been measured earlier also on a hexanuclear
iron(III) cluster, but has been found afterwards to be less reliable due to instrumental
artifacts.\cite{OW_Fe6ins,HA_private}

For an octanuclear ring with molecular $S_4$ symmetry, the position vectors are simply ${\bf R}_i = R_0 \left(
\cos(i\pi/4), \sin(i\pi/4), (-1)^i R_z \right)$, where $R_0$ is the "radius" of the ring ($R_0 = 4.427\,${\AA}
for Cr$_8$) and $R_z$ determines by how much the ions lie above or below the plane of the ring. Since $R_z \ll
R_0$, it shall be neglected here (but can be incorporated straightforwardly if felt to be required). For $R_n$
then holds $R_1 = R_7 = \sqrt{2-\sqrt{2}}R_0$, $R_2=R_6=\sqrt{2} R_0$, $R_3=R_5= \sqrt{2+\sqrt{2}}R_0$, and
$R_4=2R_0$. Also, it follows that $C^2_0({\bf R}_n) = -1/2$. This is actually valid for the weaker condition
$R^2_z \ll R^2_0$.

For finite rings with antiferromagnetic Heisenberg interactions it is well established that, as function of the
total spin quantum number $S$, the lowest lying states form a set of rotational bands.\cite{OW_spindynamics} The
energies of the states of a particular rotational band follow the Land\'e rule $E(S) \propto S(S+1)$ as for a
rigid rotator.\cite{JS_rotationalbands,OW_spindynamics} The set of bands can be distinguished into $L$- and
$E$-band: The $L$-band consist of the states for which $q$ toggles as function of $S$ between $q = 0$ and $q =
N/2$, the $E$-band embraces the lowest lying states with $q \neq 0,N/2$, which are essentially spin waves in
character.\cite{OW_spindynamics,PWA_spinwaves} The energies of the states of these bands can be approximated
excellently by $E(S,q) = {1\over 2} \Delta S(S+1) + \epsilon(q)$. $\Delta$ characterizes the "curvature" of the
bands and coincides with the energy gap between ground and first excited state. $\epsilon(q)$ measures the
"offset" of a particular band, as classified by $q$. Whereby, $\epsilon(q) \approx const |\sin(2\pi
q/N)|$.\cite{OW_spindynamics,PWA_spinwaves} Importantly, the $L$-band is well separated in energy from the
higher lying $E$-bands, i.e. $\Delta \ll \epsilon(q)$ for $q \neq 0,N/2$.

In actual INS experiments only the transitions between states within the $L$-band and those from the $L$-band
into the $E$-band are relevant since $\Delta$ is typically on the order of meV. For transitions within the
$L$-band $q-q'=N/2$ holds; and for transitions from the $L$-band into the $E$-band $q-q' \neq 0,N/2$ (values for
$q$ should always be understood modulo $N$).

First, the $L=0$ contribution in Eq.~(\ref{I_cyclic}), i.e. $f^{0}_{qq'}(Q)$ will be considered. For Cr$_8$,
$q-q'=4$ for transitions between states of the $L$-band, directly yielding

\begin{subequations}
\label{f0_cyclic}
\begin{eqnarray}
\label{f04_cyclic}
 f^0_{0,4} &=& 1 + 2 \sum_{n=1}^3 (-1)^n j_0(QR_n) + j_0(QR_4)
\end{eqnarray}

for the interference term. For transitions from a state of the $L$-band to states of the $E$-band one
additionally has to take into account the degeneracy of the states with $q$ and $N-q$ due to the overall $D_N$
symmetry. However, inspection of Eq.~(\ref{f_cyclic}) shows that the interference terms are equivalent for $q'$
and $N-q'$ (and of course analogous for $q$). One finds

\begin{eqnarray}
f^0_{0,1} &=& 1 + \sqrt{2} \left[j_0(QR_1) - j_0(QR_3)\right] - j_0(QR_4) \\
f^0_{0,2} &=& 1 - 2 j_0(QR_2) + j_0(QR_4) \\
f^0_{0,3} &=& 1 - \sqrt{2} \left[j_0(QR_1) - j_0(QR_3)\right] - j_0(QR_4).
\end{eqnarray}
\end{subequations}

\begin{figure}
\includegraphics{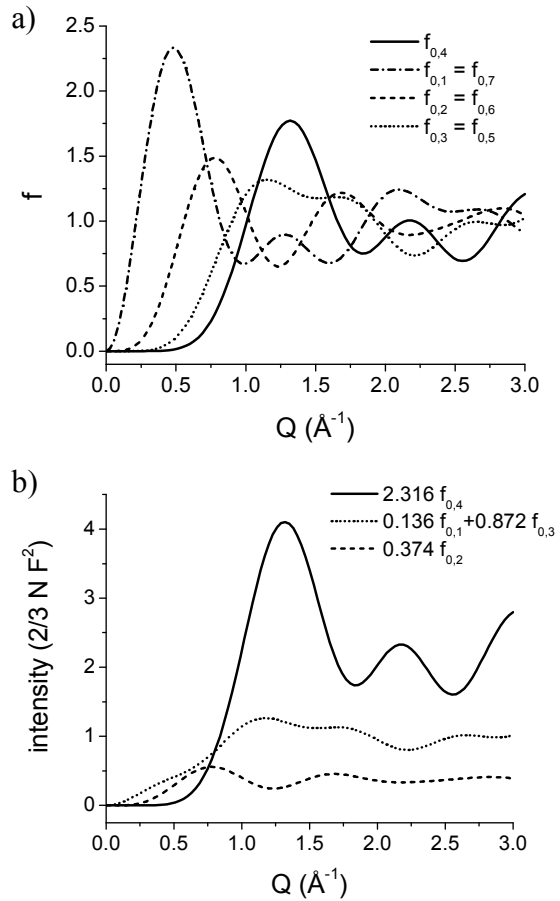}
\label{fig1} \caption{(a) $Q$-dependence of the interference terms $f^0_{0,4}$, $f^0_{0,1}$, $f^0_{0,2}$ and
$f^0_{0,3}$ as given in Eq.~(\ref{f0_cyclic}) for an octanuclear cyclic ring. (b) Dependence of the inelastic
neutron scattering intensity as function of momentum transfer $Q$ for the three observable transitions at low
temperatures of an octanuclear antiferromagnetic Heisenberg ring with spin-3/2 (see text). For both panels $R_0
= 4.427$~{\AA} was used as appropriate for the cyclic molecular cluster Cr$_8$.}
\end{figure}

These functions are plotted in Fig.~1(a) with $R_0$ chosen as appropriate for the Cr$_8$ cluster. To compare
with experiment one additionally has to consider the quasi degeneracy of the spin levels with $q=1,7$ and
$q=3,5$ because of the $|\sin(2\pi q/N)|$ like dependence of $\epsilon(q)$ on $q$. Since this quasi degeneracy
cannot be resolved experimentally, one should sum these contributions for a comparison with observable peaks.
This requires a knowledge of the oscillator strengths ${\bf \tilde{S}}^2_1 = \sum_{\alpha} \langle \tau q
|S_{1\alpha}| \tau' q'\rangle^2$ for each of the four contributing transitions. Due to symmetry reasons the
oscillator strengths for $q'$ and $N-q'$ are equivalent. The dependence on $q$ is approximated by
$\sqrt{1-\cos(q2\pi/N)/1+\cos(q2\pi/N)}$,\cite{GM_sumrules} but for precise results the oscillator strengths
should be calculated, e.g. numerically. For the possible transitions starting at the $S=0$ ground state, they
are given in Table~\ref{tab1}.

Combining Eqs.~(\ref{I_cyclic}) and (\ref{f0_cyclic}) and using Table~\ref{tab1}, the $Q$-dependence of the INS
intensity for the three transitions as observable for Cr$_8$ at low temperatures can be expressed analytically
and is shown in Fig.~1(b) in units of $2/3NF^2(Q)$. These curves were found to be in excellent agreement with
experiment.\cite{OW_Cr8ins}

Since $C^2_0({\bf R}_n) = -1/2$, the calculation of $f^{2}_{qq'}(Q)$ proceeds in exactly the same way as for
$f^{0}_{qq'}(Q)$. One obtains basically Eqs.~(\ref{f0_cyclic}) multiplied by $-1/2$ and with $j_0(QR_n)$
replaced by $j_2(QR_n)$. Thus, the magnetic anisotropy in the cyclic spin clusters mixes in some $j_2$
contributions, but since the anisotropy in general is weak compared to the antiferromagnetic coupling, the
$Q$-dependence of the INS intensity is rather well described by $f^{0}_{qq'}(Q)$ alone.

\begin{table}
\caption{\label{tab1} ${\bf \tilde{S}}^2_1 = \Sigma_\alpha \langle \tau q |S_{1\alpha}| \tau' q'\rangle^2$ for
the possible transitions from the ground state ($S=0, q=0$) to the $L$-band ($S=1,q'=4$) and to the $E$-band
($S=1,q'=1,2,3,5,6,7$) for an octanuclear antiferromagnetic spin-3/2 Heisenberg ring.}
\begin{ruledtabular}
\begin{tabular}{ccccc}
q'= & 1,7 & 2,6 & 3,5 & 4  \\
\hline
 & 0.06802 & 0.1870 & 0.4359 & 2.316 \\
\end{tabular}
\end{ruledtabular}
\end{table}

%

\section{Tetranuclear starlike spin clusters}
\label{sTET}

For further illustration of the use of Eq.~(\ref{main_result}), the tetranuclear starlike clusters will be
considered.\cite{Fe4} In these clusters a central spin is surrounded by three spins which form an almost perfect
equilateral triangle. Recently, such systems incorporating four iron(III) ions were investigated and attracted
considerable interest as they were found to behave as single molecule magnets and show the still rare phenomenon
of resonant quantum tunneling.\cite{Fe4} For one member of this class of compounds the INS spectra has been
measured recently.\cite{GA_Fe4ins,GA_Fe4interins} The $Q$-dependence has not yet been measured
reliably,\cite{GA_Fe4interins} but can be expected to be investigated with better accuracy in the near future.

The actual symmetry of these molecules is less than threefold.\cite{Fe4,GA_Fe4ins} They exhibit a twofold
symmetry axis which passes through the central spin and one of the surrounding spins. The magnetic anisotropy,
accordingly, is biaxial, but much smaller than the antiferromagnetic coupling between central spin and
surrounding spins. It shall thus be neglected here. This essentially corresponds to neglecting the $L=2$ terms
in Eq.~(\ref{I}), similar to the above discussion for the cyclic clusters. Also, the geometrical deviation from
threefold symmetry is very small. It is thus a very good approximation to take the spin permutational symmetry
as $D_3$.

The arguments to calculate the interference terms are very similar to those for the cyclic spin clusters. The
new feature which arises here is that the spins divide into two classes $\gamma$. Again, the subgroup $C_3$ will
be considered. Since there is no group element $T^n$ ($n = 0,1,2$) which maps the central spin (numbered by 4
here) onto one of the surrounding spins, one obtains the two classes $\{1,2,3\}$ and $\{4\}$. It should be noted
here, that one obtains exactly the same classes even for the full symmetry group $D_3$. The $Q$-dependence is
determined by the interference terms $f^{qq'}_{11}(Q)$, $f^{qq'}_{14}(Q)$, $f^{qq'}_{41}(Q)$, and
$f^{qq'}_{44}(Q)$ in Eq.~(\ref{ggf}). In the present case, however, $f^{qq'}_{14}(Q) = f^{qq'}_{41}(Q)$ and
$f^{qq'}_{44}(Q) = F^2(Q) {\bf \tilde{S}}_4^2$. The averaged INS cross section is then

\begin{subequations}
\label{I_star}
\begin{eqnarray}
\bar{I}_{\tau q\tau'q'}(Q) = {2\over3} F^2(Q) \times\cr \left[ {\bf \tilde{S}}_1 \cdot {\bf \tilde{S}}_1
f^{qq'}_{11}(Q) + {\bf \tilde{S}}_1 \cdot {\bf \tilde{S}}_4 f^{qq'}_{14}(Q) + {\bf \tilde{S}}_4 \cdot {\bf
\tilde{S}}_4  \right]
\end{eqnarray}

with

\begin{eqnarray}
f^{qq'}_{11}(Q) &=& 3 + 6 j_0(Q R_{12}) \cos\left( {2\pi\over 3}(q-q') \right) \\
f^{qq'}_{14}(Q) &=& 3 j_0(Q R_{14}) \delta_{qq'}.
\end{eqnarray}
\end{subequations}

Similar to the cyclic clusters, spin levels with $q=0$ belong to an one-dimensional IR and that with $q=1,2$ to
the two-dimensional IR of $D_3$. The latter are thus degenerate and one should add the respective INS
intensities.

In the starlike clusters, the antiferromagnetic coupling between the spin-5/2 iron(III) ions leads to a $S=5$
ground state \cite{Fe4,GA_Fe4ins} belonging to $q=0$. The first excited states, approximately 80$\,$K above the
ground state, are made up of two $S=4$ levels \cite{Fe4,GA_Fe4ins} belonging to $q=1,2$. This excitation has
been observed in experiment. \cite{GA_Fe4interins} Its INS intensity is calculated as

\begin{eqnarray}
\label{I_Cr8_0,12}
 \bar{I}_{0,12}(Q) = {2\over3} F^2(Q) \times\cr 3 \left\{ 2 {\bf \tilde{S}}_1 \cdot {\bf
\tilde{S}}_1 \left[ 1- j_0(Q R_{12}) \right] + {\bf \tilde{S}}_4 \cdot {\bf \tilde{S}}_4  \right\}
\end{eqnarray}

since, as for the cyclic clusters, the interference terms remain unchanged by $q' \rightarrow N-q'$. Up to a
constant, the $Q$-dependence is here again determined by the spatial symmetry properties of the molecule.

Equation~(\ref{I_Cr8_0,12}) is expected to describe the $Q$-dependence of the observed $S=5 \rightarrow 2 \times
S=4$ transition well, not only because of the reasons given above, but also because the only structural element
which enters is $R_{12}$. This implies a strategy to use the $Q$-dependence as a check for effects which break
the $D_3$ spin permutational symmetry (the $Q \rightarrow 0$ argument is not working here because of the $Q$
independent contribution $f^{qq'}_{44}$). Such effects lead in particular to a contribution proportional to
$j_0(QR_{14})$ which would be absent otherwise [Eq.~(\ref{I_Cr8_0,12})]. Of course, $L=2$ contributions also
will differ, but they remain small as the magnetic ansiotropy is much smaller than the coupling.

Transitions from the ground state to even higher lying states or within the excited states should be very
difficult to observe experimentally and thus are not discussed here. But transitions within the $S=5$ ground
state multiplet, corresponding to $q=0 \rightarrow q'=0$, were observed. \cite{GA_Fe4ins} The calculation of
their $Q$-dependencies, however, is essentially just a repetition of the above considerations.

%

\section{Conclusions}
\label{sCON}

In this work, it has been shown for the two cases of the cyclic and the star like clusters how the spatial
symmetry of the molecule predetermines interference terms and thereby the $Q$-dependence of the INS intensity.
The symmetry elements of the spin permutational symmetry group $\mathcal{G}$ impose relations among the matrix
elements at different spin sites. The number of classes $\gamma$ allowed by the spin permutational symmetry
turned out to be essential: The smaller the number of classes is, the fewer spin matrix elements remain
uncorrelated, and the more effective the whole procedure becomes. The two cases discussed here are in a sense
extreme as the number of classes $\gamma$ produced by the symmetry operations was very small, being even one for
the cyclic spin cluster.

The procedure of how to calculate the effects of the spatial symmetry has been worked out here only for the case
of one-dimensional IRs. This might appear as a significant restriction, but it is sufficient for many practical
cases. For instance, in the cases presented here, the full symmetry group actually had two-dimensional IRs. This
has been handled by first considering a subgroup with only one-dimensional IRs, calculating the interference
terms for this situation, and then, in a second step, to take into account the degeneracies imposed by the full
symmetry by simply adding the corresponding intensities. In the present cases this was quite trivial as the
interference terms were identical for the involved degenerate states. In the general case, one has to work out
the relationship between the involved matrix elements. However, since it is entirely determined by
transformation properties it can be calculated using the generalized Wigner-Eckhard theorem and the irreducible
operators as appropriate for the symmetry group $\mathcal{G}$.\cite{AM_grouptheory}

This approach worked for the cyclic and star like clusters without loss of information about the $Q$-dependence
because the symmetry elements of the subgroup were complete enough to produce the minimal number of classes.
That is, exactly the same class structure as for the full symmetry group was obtained. Certainly, if this does
not hold, one either has to accept additional undetermined matrix elements in the analytical expression for the
$Q$-dependence, or has to start right away from the general, but somewhat involved, equation
Eq.~(\ref{Sj_to_Si}).

It is noteworthy that the outlined strategy works for quite a number of important cases. For instance,
tetranuclear clusters of the so called cubane structure are very frequent and constitutes an important class of
tetranuclear compounds. Actually, a significant number of tetranuclear clusters with a different topology (e.g.
grids, squares, or chains) became available only recently due to advances in inorganic chemistry. The metal
centers in these complexes approximate a tetrahedron, i.e. exhibit a rather high molecular symmetry. However,
the C$_4$ subgroup is sufficient to connect all spin centers by symmetry elements resulting in only one class
$\gamma$ including all spin centers. Thus, these clusters can be handled in exactly the same way as the cyclic
spin clusters.

In many cases, the molecular cluster does not exhibit a high symmetry exactly, but approximates it with small
distortions. The Bessel functions will be affected only slightly since deviations from the regular positions of
the spin centers are smaller than the distances between the spin centers. Furthermore, the distorted structure
also leads only to small changes in the parameters of the spin Hamiltonian, which typically have no effect on
the energy spectrum up to first order in perturbation theory.\cite{OW_CsFe8} As a result, the interference terms
are actually rather insensitive to weak distortions from an assumed optimal symmetry. The $Q$-dependencies as
calculated for the approximate symmetry are then nonetheless useful, in particular as the accuracy of
experimental $Q$-dependencies is often limited.

With regard to the behavior of the INS intensity for $Q$ approaching zero, it has been found for the cyclic
clusters that it drops to zero if states with different spatial spin quantum numbers are involved in the
transition. Vice versa, it will approach a finite value only if the spatial spin quantum numbers coincide. This
is a very general feature for clusters for which all spin centers are connected by elements of the symmetry
group $\mathcal{G}$, i.e. for which only one class $\gamma$ appears. With this reasoning in mind, the different
behavior of the $Q$-dependencies of the transitions observed for e.g. the dimer Tb$_2$Br$_9^{3-}$ becomes
immediately transparent \cite{AF_Tb2ins} - it is a result of the inversion symmetry of the dimer. These
differences for $Q\rightarrow0$ were used in this case with advantage to confirm the assignment of observed
peaks to theoretically expected transitions.\cite{AF_Tb2ins} As is clear from the discussion of the tetranuclear
star like cluster, this kind of argumentation does not apply to cases in which more than one class $\gamma$
arise.

%

\begin{acknowledgments}
OW thanks the Deutsche Forschungsgemeinschaft for partial financial support.
\end{acknowledgments}

%

%
\end{document}